\documentclass[twocolumn,runningheads]{svjour2}
\smartqed  
\usepackage{graphicx}
\usepackage{amssymb}
\journalname{Astrophysics and Space Science}
\begin{document}

\title{Gamma ray signatures of ultra high energy cosmic ray accelerators: electromagnetic cascade versus synchrotron radiation of secondary electrons
}


\author{Stefano Gabici         \and
        Felix A. Aharonian 
}


\institute{S. Gabici and F. A. Aharonian \at
              Max-Planck-Institut f\"ur Kernphysik, 
	      Heidelberg, Germany 
}

\date{Received: date / Accepted: date}

\maketitle

\begin{abstract}
We discuss the possibility of observing ultra high energy cosmic ray sources in high energy gamma rays. Protons propagating away from their accelerators produce secondary electrons during interactions with cosmic microwave background photons. These electrons start an electromagnetic cascade that results in a broad band gamma ray emission. We show that in a magnetized Universe ($B \gtrsim 10^{-12}$ G) such emission is likely to be too extended to be detected above the diffuse background. A more promising possibility comes from the detection of synchrotron photons from the extremely energetic secondary electrons.
Although this emission is produced in a rather extended region of size $\sim 10 Mpc$, it is expected to be point-like and detectable at GeV energies if the intergalactic magnetic field is at the nanogauss level.
\PACS{98.70.Sa \and 98.70.Rz}
\end{abstract}


Understanding the origin of the Ultra High Energy Cosmic Rays (UHECR) with energy above $\sim 5 \; 10^{19}$ eV 
constitute a real challenge for theoretical models, since their acceleration requires extreme conditions hardly fulfilled by known astrophysical objects.
Previous studies considered a number of potential sources, including gamma ray bursts, active galactic nuclei, large scale jets and neutron stars, but results are still inconclusive \cite{reviews}.  

If, as many believe, UHECRs have an extragalactic origin, a suppression, known as the Greisen-Zatsepin-Kuzmin (GZK) cutoff, should be observed in the spectrum at an energy 
$\sim 7 \; 10^{19} {\rm eV}$ 
\cite{GZK}. 
This is because during propagation energetic protons undergo $p\gamma$ interactions (mainly photo-pion production) in the 
Cosmic Microwave Background (CMB) 
with an energy loss length which decreases dramatically from $\sim 650$ to $\sim 20 \, {\rm Mpc}$ between energies of $7 \; 10^{19}$ to $3 \; 10^{20} {\rm eV}$ \cite{pgamma}. 
The detection of a number of events above $7 \, 10^{19}$ eV 
raised interest in non-acceleration 
models 
that
involve new physics and do not require the existence of a cutoff in the spectrum \cite{reviews}.
However, the issue of the presence 
of the GZK feature 
is still debated \cite{daniel}. 
In the following we assume that astrophysical objects capable of accelerating UHECRs 
do exist and we discuss possible ways to identify them. 

If UHECRs are 
not significantly 
deflected by the 
intergalactic magnetic field (IGMF) 
their arrival direction should point back to the position of their accelerators.
In principle, this fact can 
be used to identify accelerators located at distances smaller than the proton loss length, that constitutes a sort of horizon for CRs. However, even in this case, the small event statistics and the high uncertainty 
in the determination of the CR arrival direction 
would make any identification problematic. 
The better angular resolution and 
collection area of AUGER will hopefully provide 
adequate data to address this issue.


The interactions between UHECRs and CMB photons generate secondary gamma rays and electron-positron pairs (hereafter referred to as {\it electrons}) that in turn initiate an electromagnetic (EM) cascade in the universal photon background \cite{cascade}.
Such a cascade would appear to a distant observer as a flux of GeV/TeV photons.
Recently, Ferrigno et al. \cite{carlo} suggested to 
identify the sources of UHECRs 
by searching for this radiation.
Their calculations show that, in an unmagnetized Universe ($B_{IGMF} = 0$ G), a steady source emitting isotropically $2 \; 10^{43}$ erg/s in form of UHECRs ($E > 10^{19}$ eV) can be detected by a Cherenkov telescope like HESS up to a distance of $\sim 100$ Mpc.
In this 
case the source would be point-like, because in the absence of IGMF the EM cascade is one-dimensional and propagates radially away from the accelerator. 
Unfortunately, the scenario changes dramatically if 
the more realistic case of a 
magnetized Universe is considered. This is because low energy electrons produced during the last steps of the cascade are effectively deflected and eventually isotropized if their Larmor radius is smaller or comparable with the Compton cooling length. This condition is satisfied when the IGMF is above
$
B_{iso} \sim 10^{-12} (E_{\gamma}/TeV) {\rm G}
$,
$E_{\gamma}$ being the energy of the Compton photon. This implies that \textit{unless the IGMF is extremely weak} ($<< B_{iso}$), 
electrons emit Compton photons after being fully isotropized. Thus, \textit{an extended halo of emitting pairs forms around UHECR sources}. The radiation from the halo is emitted isotropically, resulting in a \textit{very extended, and thus hard to be detected, gamma ray source} \cite{cascade}.

However, if the IGMF close to the accelerator is at the ${\rm nG}$ level, first generation electrons ($E \sim 10^{19}$ eV) generated during $p\gamma$ interactions cool rapidly by emitting ${\rm GeV}$ synchrotron photons and the development of the cascade is strongly inhibited. 
We propose the possibility to detect these synchrotron photons from UHECR sources \cite{noi}.  
This is of great interest because of the following reasons.
Both synchrotron emitting electrons and parent protons are extremely energetic and not appreciably deflected by the IGMF, at least on the first Mpcs distance scale.
For this reason, \textit{synchrotron photons are emitted in the same direction of parent protons}. Thus, they move away from the source almost radially, and \textit{the observed radiation is expected to be point-like}, and thus easily detectable and distinguishable from the extended cascade component.
Remarkably, a detection of these sources would allow to infer the value of the IGMF close to the accelerator, constraining it in the range $\approx 10^{-10} \div 10^{-8}$ G. 
Finally, since the Universe is transparent to ${\rm GeV}$ photons, powerful UHECR accelerators located outside the CR horizon might be identified in this way.  

\section{Development of the electromagnetic cascade}

\begin{figure}
\centering
\includegraphics[width=0.4\textwidth]{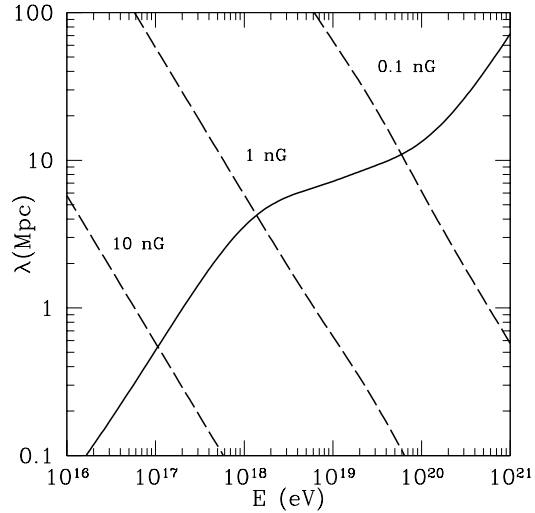}
\caption{\label{eloss}Effective electron loss length. Solid: Compton/pair-production losses in the CMB and radio background (with a cutoff at $2 {\rm MHz}$ \cite{radiobackground}). Dashed: synchrotron losses. 
}
\end{figure}

In this section we describe the development of the EM cascade initiated by a UHECR.
Consider a proton with energy $E_{p,20} = E_p/10^{20}$ eV. The typical energies of photons and electrons produced in $p\gamma$ interactions are $\sim 10^{19} E_{p,20}$ and $\sim 5 \; 10^{18} E_{p,20}$ eV respectively \cite{pgamma}. 
In the absence of IGMF, such electrons and photons interact via Compton and pair production processes with photons in the CMB and radio background.
In general, this would lead to the development of an EM cascade, in which the number of electrons and photons increases rapidly.
In fact, due to the extremely high energy of the particles considered here, each interaction occurs in the limit $\Gamma = \epsilon_b E \gg 1$, where $\epsilon_b$ and $E$ are the energies of the background photon and of the energetic electron (photon) respectively, both calculated in units of the electron rest mass energy.
Under this condition the Compton scattering is in the extreme Klein-Nishina limit, namely, the upscattered photon carries away most of the energy of the incoming electron.
The same happens during a pair production event, in which most of the energy goes to one of the two outgoing electrons.
Therefore the problem reduces essentially to a single-particle problem, in which a leading particle loses continuously energy and changes state from electron to photon and back 
due to alternate Compton/pair production interactions.
Thus, the effective loss length of an energetic electron can be identified with the loss length of the leading particle \cite{gould}. 
When the leading particle loses its energy until $\Gamma \approx 1$, 
the cascade enters the particle multiplication phase, in which the particle energy is roughly divided in half in every collision. This phase ends up with a large number of low energy electrons and photons.
On the other hand, if a IGMF is present, electrons also lose energy via synchrotron emission, subtracting energy to the cascade. 
In Fig.~\ref{eloss} we show the effective electron loss length for Compton/pair production (solid), together with the synchrotron loss length for a IGMF equal to $0.1$, $1$ and $10 ~ {\rm nG}$ (dashed lines).
It can be seen that if the IGMF is at the level of 1 nG or more, 
all the electrons 
with energy above $\sim 10^{18}$ eV cool fast via synchrotron losses and the development of the cascade is strongly suppressed.

Summarizing, three different regimes, corresponding to different values of the IGMF, can be distinguished:

\begin{itemize}
\item[$\bullet$]{{\bf Regime I:} $B_{IGMF} \ll B_{iso} \sim 10^{-12}$ G. The EM cascade is not affected at all by the 
IGMF.}
\item[$\bullet$]{{\bf Regime II:} $B_{iso} \le B_{IGMF} \ll B_{syn} \sim 10^{-9}$ G. No energy is subtracted to the EM cascade due to synchrotron losses, but low energy electrons are effectively isotropized by the IGMF.}
\item[$\bullet$]{{\bf Regime III:} $B_{IGMF} \ge B_{syn}$. The development of the EM cascade is strongly suppressed since its very first steps due to strong synchrotron losses.}
\end{itemize}

\section{Regime I: one-dimensional cascade}

If the IGMF strength is much less than $B_{iso} \sim 10^{-12}$ G, electrons in the cascade do not suffer synchrotron losses, nor are they deflected. Thus, the EM cascade develops along a straight line. In this case, the calculations by Ferrigno et al. apply, and nearby and powerful UHECR sources might be detected as point like TeV sources by currently operational Cherenkov telescopes \cite{carlo}.

In principle, since the strength of the large scale IGMF is basically unknown \cite{Bfield}, such a low values of the field cannot be ruled out. However, this is probably not a good assumption in the vicinity of UHECR accelerators, where the IGMF is expected to be appreciable, especially if such accelerators are, as it seems reasonable to believe, correlated with the structures in the Universe. 
The cascade might still be one-dimensional if its last steps develop sufficiently far away from the source, in a region of very low IGMF. Another necessary condition is that the IGMF close to the source must be small enough ($\ll 10^{-9}$ G) to avoid a suppression of the cascade due to synchrotron losses of first generation electrons. 

\section{Regime II: extended pair halos}

If the IGMF is strong enough to deflect the electrons in the cascade, but not enough to make synchrotron losses relevant (namely, $10^{-12} {\rm G} \le B_{IGMF} \ll 10^{-9}$ G), then the EM cascade fully develops, low energy electrons are isotropized, and a very extended pair halo forms around the UHECR source. For an isotropic source, the size of the halo can be roughly estimated as follows. Let $E_{\gamma}^{obs}$ be the energy of the gamma ray photons observed from the Earth. Such photons are CMB photons Compton-upscattered by electrons with energy $E_e \sim 20 (E_{\gamma}^{obs}/{\rm TeV})^{1/2}$ TeV. These are the electrons forming the pair halo. 
Since electrons are rapidly isotropized in the IGMF, one can assume that they do not propagate away from the sites in which they are created.
Electrons in the halo are in turn produced by parent photons with energy $E_{\gamma}^{par} \lesssim E_e$. 
Since the photon mean free path against pair production in the infrared background $\lambda_{pp}$ decreases rapidly with increasing energy \cite{IRabs}, we can safely neglect the contribution to the halo size from older generation (higher energy) photons.  
Thus, the size of the halo 
can be roughly estimated as $l_{halo} \sim \lambda_{pp}(E_{\gamma}^{par})$ \cite{cascade}.
For a $\sim 20$ TeV photon the mean free path is about a few tens of megaparsecs (see Fig. 2 in \cite{felixICRC}). In fact, for the situation considered here, the size of the halo is even bigger, since the UHECR protons and the first generation electrons 
propagate $\sim 10 \div 20$ Mpc before 
initiating the EM cascade (see Fig. 2).
    Thus, a conservative estimate of the apparent angular size of the halo at 1 TeV can be given by: $\vartheta \sim (l_{halo}/D) \approx 10^o (l_{halo}/20Mpc) (D/100Mpc)^{-1}$, where $D$ is the distance of the source.
This indicates that \textit{pair halos are extremely extended}, bigger than the field of view of Cherenkov telescopes (the HESS field of view is $\sim 5^o$) \textit{and thus hardly detectable}. The problem becomes even worse if one considers also protons with energy below $10^{20}$ eV, which have much longer loss length, increasing up to $\sim 1$ Gpc for proton energies equal to $\sim 5 \; 10^{19}$ eV. On the other hand, such protons are likely to contribute only to the TeV flux of very distant sources.

In the recent work by Armengaud et al. \cite{sigl} the deflection of electrons in the IGMF has been neglected, even when a IGMF stronger than $B_{iso} \sim 10^{-12}$ G was assumed. On the other hand, the authors considered the deflection of $\sim 10^{20}$ eV protons, which is in fact totally negligible if compared with the full isotropization of electrons. 
Therefore their claim
 about the detectability of cascade gamma-rays 
seems to us over-optimistic.

The cascade emission peaks at TeV energies \cite{carlo}, making a detection by GLAST problematic.

\section{Regime III: synchrotron gamma rays}

If the IGMF close to the UHECR accelerator is at the level of 1 nG or above, the development of the cascade is strongly suppressed, since the very high energy electrons produced during $p\gamma$ interactions cool rapidly via synchrotron losses before undergoing Compton scattering. It is evident from Fig. 2 that for a $\sim nG$ IGMF this is true for electron energies well in excess of $E_e \sim 10^{18}$ eV. Such electrons emit synchrotron photons with energy 
$E_{syn} \approx 2 \, (B/{\rm nG}) (E/10^{19}{\rm eV})^2 {\rm GeV}$, detectable by GLAST.
It is important to stress that our results are sensitive only to the value of the IGMF {\it close} to the source, while they are unaffected by the value of the field on much larger scales. 
This is because synchrotron emitting electrons are produced within a proton interaction length $l_{p\gamma} \approx 10$ Mpc from the accelerator.
As a consequence, the only assumption required
is that the size of the magnetized region surrounding the accelerator must be greater or comparable with $l_{p\gamma}$.
Superclusters of galaxies constitute an example of large and magnetized regions satisfying our requirement \cite{Bfield}. 

We now estimate the angular size of the synchrotron emission.
After propagating over an interaction length, a proton of energy $E_p$ is deflected by an angle
$\vartheta_p \approx 0.8^o (10^{20}{\rm eV}/E_p) (B/{\rm nG}) \sqrt{(l_{p\gamma}/10{\rm Mpc})} \sqrt{(l_c/{\rm Mpc})}$  
where $l_c$ is the IGMF coherence length \cite{deflection}.
Due to the high energies considered, secondary electrons produced in $p\gamma$ interactions move in the same direction of the parent protons.
In a cooling time electrons are deflected by an angle $\vartheta_e \sim \alpha \lambda / R_L$,
$R_L$ being the electron Larmor radius and $\alpha$ a number of order unity representing the probability that the leading particle is actually an electron \cite{gould}.
Remarkably, if expressed as a function of the synchrotron photon energy,  
the deflection angle is independent on the magnetic field strength and reads: $\vartheta_e \approx 0.5^o (E_{syn}/10 GeV)^{-1}$.
Thus, an observer at a distance $D$ would see a source with angular size: 
$\vartheta_{obs} \approx \sqrt{\vartheta_p^2+\vartheta_e^2} \left(\frac{l_{p\gamma}}{D}\right)$ that, for $D = 100$ Mpc and for photon energies of $1 \div 10$ GeV is of the order of a fraction of a degree. 
This is comparable with the angular resolution of GLAST, that would classify these sources as point-like if they are
located at a distance of $\sim 100 \; {\rm Mpc}$ or more. 
This leads to the important conclusion that, \textit{even if synchrotron photons are produced in an extended region of size $\sim l_{p\gamma}$ surrounding the accelerator, the resulting gamma ray source would appear point-like to a distant observer}.

If the proton spectrum extends well above $10^{20}$ eV, these sources might also be detected by 
Imaging Atmospheric Cherenkov Telescope arrays (IACT) operating at energies above $100 \; {\rm GeV}$. The angular resolution of these instruments is a few arcminutes, and thus the sources will appear extended. However, it has been proven that IACT are powerful instruments to image extended gamma ray sources \cite{SNR}, and thus they might still detect and map the emission from UHECR sources. Finally, powerful UHECR accelerators located at a distance of 1 Gpc or more would appear as point sources. In the next section we discuss the energy requirement for a detection.

\begin{figure}
\includegraphics[width=0.4\textwidth]{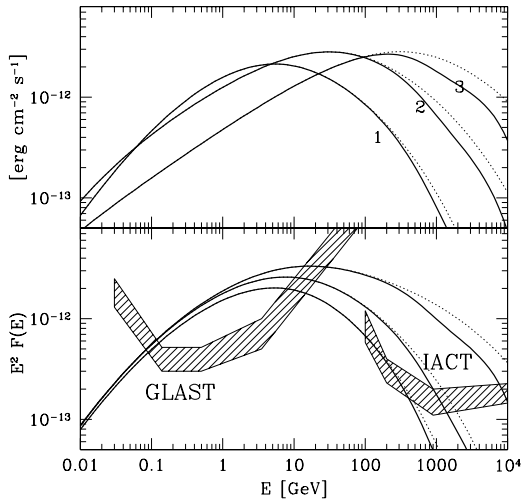}
\caption{\label{spectra}Spectra for a source 
located at 
$100{\rm Mpc}$.
The luminosity in UHECRs is $2 \; 10^{44}{\rm erg/s}$, with spectral index is $\delta = 2$. TOP: $B_{IGMF} = 0.5$(curve 1), $5$ (2), $50{\rm nG}$(3), $E_{cut} = 10^{21} {\rm eV}$. BOTTOM: $E_{cut} = 5 \, 10^{20}$, $10^{21}$, $5 \, 10^{21} {\rm eV}$, $B_{IGMF} = 1{\rm nG}$. Dotted: intrinsic spectra. Solid: spectra after absorption.}
\end{figure}

\section{Detectability and energetics}

Fig.~\ref{spectra} shows synchrotron spectra 
for a steady source
at a distance of $100 ~ {\rm Mpc}$ in a uniformly magnetized region of size 20 Mpc.
Since the radiation is produced in a region about one interaction length away from the source, the actual structure of the field is not crucial.
Steady state proton and electron spectra have been calculated taking into account all the relevant energy losses and proton escape from the magnetized region. 
Solid lines have been computed 
taking into account the opacity of the Universe to very high energy photons due to pair production in the cosmic infrared background \cite{IRabs},
while dotted lines show the unabsorbed spectra.
The total luminosity in UHECRs with energy above $10^{19} {\rm eV}$ is $L_{UHE} = 2 ~ 10^{44} {\rm erg/s}$, with a differential energy distribution $\propto E^{-\delta} exp(-E/E_{cut})$, with 
$\delta =2$. 
Results are quite insensitive to the slope of the spectrum.

In the top panel of Fig.~\ref{spectra} $E_{cut} = 10^{21} {\rm eV}$ 
and the IGMF 
is equal to $0.5$, $5$ and $50 \; {\rm nG}$ (curves $1$, $2$ and $3$).
If the IGMF is significantly greater than $\sim 50 \; {\rm nG}$, the peak of the emission falls at TeV energies, where absorption is very strong.
On the other hand, if the field is well below $\sim 0.5 \; {\rm nG}$, synchrotron emission becomes unimportant and the cascade contribution dominates.
However, \textit{for the broad interval of values of the IGMF strength between $0.5$ and $50 \; {\rm nG}$, the formation of a synchrotron point-like gamma ray source seems to be unavoidable}. 

In the bottom panel of Fig.~\ref{spectra}, our predictions are compared with the sensitivities of GLAST and of a generic 
IACT such as HESS or VERITAS. 
A IGMF of $1 \; {\rm nG}$ is assumed and the different curves refer to values of the cutoff energy in the proton spectrum equal to $5 \; 10^{20}$, $10^{21}$ and  $5 \; 10^{21} {\rm eV}$ (top to bottom).
For such a field, the condition for the detectability of a point source by GLAST is roughly $L_{UHE} \ge 8 \; 10^{43} \div 2 \; 10^{44} (D/100{\rm Mpc})^2 {\rm erg/s}$ for $\delta = 2.0 \div 2.6$.
In contrast, for IACT arrays the minimum detectable luminosity is roughly 2 orders of magnitude higher, since the source has to be located at a distance of $\sim 1 {\rm Gpc}$ in order to appear point-like. However, less powerful accelerators can still be detected as extended sources.
Since the peak of the emission falls at $\sim 10 ~ {\rm GeV}$, future IACT operating in the energy range $10 \div 100$ GeV would be powerful tools to search for these sources.  

If the CR spectrum smoothly 
extends down to GeV energies with slope $\delta = 2$, the required total CR luminosity for a source to be detected by GLAST is $L_{CR} \ge 5 \, 10^{44} (D/100{\rm Mpc})^2 {\rm erg/s}$.
For a beamed source, the required luminosity is reduced by a factor $f_b \sim 0.02 (\vartheta_b/10^o)$, $\vartheta_b$ being the beaming angle,
and the detectability condition reads: $L_{CR} \ge 10^{43} (f_b/0.02) (D/100{\rm Mpc})^2 {\rm erg/s}$.
This luminosity is small if compared, for example, with the 
power of an AGN jet, that can be as high as 
$10^{47} {\rm erg/s}$ \cite{ghisella}.
Thus, 
astrophysical objects that can in principle satisfy the energy requirement for a detection do exist.

\end{document}